# Optically pumped GeSn micro-disks with 16 % Sn lasing at 3.1 µm up to 180K

V. Reboud[1*], A. Gassenq[2], N. Pauc[2], J. Aubin[1], L. Milord[1], Q. M. Thai[2], M. Bertrand[1], K.Guilloy[2], D. Rouchon[1], J. Rothman[1], T. Zabel[3], F. Armand Pilon[3], H. Sigg[3], A. Chelnokov[1], J.M. Hartmann[1], V. Calvo[2]

[1]*Univ. Grenoble Alpes, CEA-LETI, Minatec, 17 rue des Martyrs, 38000, Grenoble, France*

[2]*Univ. Grenoble Alpes, CEA-INAC, 17 rue des Martyrs, 38000, Grenoble, France*

[3]*Laboratory for Micro- and Nanotechnology, Paul Scherrer Institute, 5232, Villigen, Switzerland*

**ABSTRACT**

Recent demonstrations of optically pumped lasers based on GeSn alloys put forward the prospect of efficient laser sources monolithically integrated on a Si photonic platform. For instance, GeSn layers with 12.5% of Sn were reported to lase at 2.5 µm wavelength up to 130 K. In this work, we report a longer emitted wavelength and a significant improvement in lasing temperature. The improvements resulted from the use of higher Sn content GeSn layers of optimized crystalline quality, grown on graded Sn content buffers using Reduced Pressure CVD. The fabricated GeSn micro-disks with 13% and 16% of Sn showed lasing operation at 2.6 µm and 3.1 µm wavelengths, respectively. For the longest wavelength (i.e 3.1 µm), lasing was demonstrated up to 180 K, with a threshold of 377 kW/cm² at 25 K.

* vincent.reboud@cea.fr

Germanium tin (GeSn) is a Complementary Metal Oxide Semiconductor (CMOS)-compatible group IV material which can exhibit a direct bandgap [1,2], leading to potential applications in photonics [3] and microelectronics [4]. Recent progresses in Chemical Vapor Deposition (CVD) [5] allowed the demonstration of the first GeSn laser in 2015 [3]. Later on, high Sn content $Ge_{1-x}Sn_x$ layers with 8.5% to 12.5% of Sn lasing at wavelengths up to 2.5 µm and at temperatures up to 130 K were demonstrated by the same group [6]. The lasing operation was confirmed in a double heterostructure configuration using a strained Ge cap layer [7]. However, for such group-IV lasers to become an alternative to a complex integration of III-V lasers on Silicon (Si), their operating temperature is to be substantially increased.

To increase the operating temperature of group-IV semiconductor lasers, two paths are considered promising and are being currently investigated: : GeSiSn alloys [8,9] and tensile strain [10,11,12]. However, a straightforward improvement of the GeSn material itself should not be neglected: a higher Sn content in GeSn layers increases the energy difference between the L and the Γ valley [5,9,13] and shifts the emitted wavelengths toward the Mid-Infrared (MIR) [9,14]. The wavelength range between 3 µm and 3.5 µm is indeed interesting for sensing applications. There, the atmosphere transparency window overlaps with absorbing lines of various gases [15,16]. Similarly to the short wave IR (SWIR), standalone III-V lasers have been developed for the MIR [17,18,19], and they are not compatible with CMOS foundries. Therefore, GeSn materials are promising candidates for group IV laser sources fully integrated in SWIR and MIR Si photonic systems, both for sensing applications and on-chip short range optical interconnects [20,21].

The challenge in increasing the Sn content in GeSn is to preserve the crystallographic quality of the material. In this work, we demonstrate lasing operation up to 180K with optically pumped 16%

Sn GeSn microdisk cavities emitting at 3.1 µm. We attribute the significantly higher lasing temperature to (i) a stronger band offset between the L- and the Γ- valleys stemming from a higher Sn content, (ii) a better crystalline quality of the GeSn layers grown on top of step-graded GeSn buffers and (iii) the carrier confinement inside the lowest bandgap layer, preventing recombination in the more defective GeSn bottom layers [22].

Thick GeSn layers (well above the critical thickness for plastic relaxation) were grown in a 200 mm Epi Centura 5200 Reduced Pressure-Chemical Vapor Deposition (RP-CVD) cluster tool from Applied Materials[14,23,24]. Digermane ($Ge_2H_6$) and tin-tetrachloride ($SnCl_4$) were used as low temperature precursors. All growth processes, including those of the step-graded GeSn buffers, were carried out on top of Ge Strain Relaxed Buffers (SRBs) [22], reducing the lattice mismatch to the Si(001) substrates. Four configurations have been investigated: (i) a 450 nm thick GeSn active layer with nominally 13% of Sn (**sample A**), (ii) a ~ 450 nm thick GeSn active layer with nominally 16% of Sn (**sample B**), (iii) a ~ 200 nm thick GeSn 13% active layer on top of a GeSn step-graded buffer made of GeSn 8% and 9.5% layers (~ 100 nm thick each) (**sample C**), (iv) a ~ 200 nm thick GeSn 16% active layer grown on a GeSn step-graded buffer made of GeSn 8%, 9.5% and 13% layers (~ 100 nm thick each) (**sample D**). Sn incorporation in the GeSn layers was controlled by changing the growth temperature [14,23,25]. In the case of $Ge_{0.87}Sn_{0.13}$ layers (sample A), 95 % of the 200 mm wafer area presents smooth/mirror-like surface under grazing light with no significant Sn segregation. For $Ge_{0.84}Sn_{0.16}$ layers (sample B), the useful area reduces to around 30% of the 200 mm wafer surface. In the case of the GeSn layers grown on step-graded buffers, the layers were mirror-like over the whole wafer surface. Using a GeSn step-graded structure improved a lot the quality of the stacks, especially for very high Sn concentrations (i.e. 16%) [22]. All optical characterizations were performed on mirror-like areas.

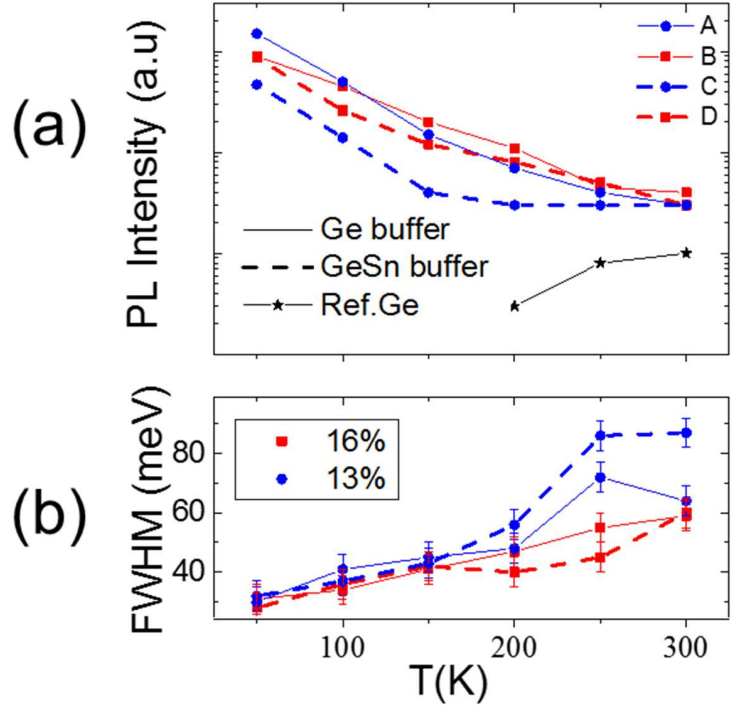

Figure 1: (a) Integrated PL intensity and (b) Full Width at Half Maximum of the PL peaks as a function of the measurement temperature for samples A to D (Table I) together with the integrated PL intensity of a Ge SRB grown on Si in the same CVD reactor.

X-ray diffraction (XRD) measurements indicate that the top GeSn layers are partially relaxed (see Supplementary Information). The residual compressive strain and the Sn content of the top layers (see Table I) were extracted from Omega-2Theta scans around the (004) XRD order and reciprocal space maps (RSM) around the (224) order (Figure S1) taking into account the small positive deviation from a straightforward interpolation between the lattice parameter of pure Ge and Sn [26]. The GeSn top layers were slightly less relaxed and the residual compressive strain higher for step-graded samples (samples C and D) than for their SRB counterparts (samples A and B). To fully relax the residual strain, suspended micro-disks were fabricated in those layers [6,27]. Strain values at the edges of suspended micro-disks were measured by Raman spectroscopy (Figure S2,

Supplementary Information). The residual strain and the Sn content in the GeSn layers laying on top of the four samples are provided in Table I for unsuspended layers (before elastic strain relaxation) and for suspended micro-disks (i.e after strain relaxation). The theoretical bandgap was determined using the conventional deformation potential theory [28,29], with a linear interpolation between the deformation potential of pure Ge and pure Sn [3,9] and a bowing parameter at 2.4 eV found in reference [12]. The theoretical bandgaps for the L ($Eg_L$) and Γ valleys ($Eg_\Gamma$) at 300 K are added in Table II of the Supplementary Information. A good agreement is found with the measured Γ bandgaps ($Eg\Gamma$) at 300 K provided in Table II. The larger compressive strain in samples C and D using a step-graded buffer results in a smaller energy difference between the L and the Γ bandgap, of only 15meV in sample C compared to 30 meV for sample A and 53meV (sample D) compared to 63meV (sample B), respectively. However, larger Sn content of the samples D and B compared to samples C and A, favorably increases energy differences between the L and the Γ bandgap, to compare with $k_BT$ = 25 meV at room temperature. During the fabrication of micro-disks, the residual compressive strain is released, increasing the difference between the L and the Γ bandgap to around 90 meV for micro-disks with 13% Sn and to around 150 meV for micro-disks with 16% Sn (Table I).

| Micro-disks fabricated on | | | Before strain relaxation | | After strain relaxation | |
|---|---|---|---|---|---|---|
| Sample | Growth buffer | $x_{Sn}$ | $\varepsilon_{//}$ | $Eg_L$-$Eg_\Gamma$ | $\varepsilon_{//}$ | $Eg_L$-$Eg_\Gamma$ |
| | | % | % | eV | % | eV |
| A | Ge SRB | 13.07 | -0.56 | 0.030 | ~ 0* | 0.095 |
| B | Ge SRB | 16.41 | -0.76 | 0.063 | ~ 0* | 0.152 |
| C | Step graded GeSn | 12.87 | -0.66 | 0.015 | ~ 0* | 0.092 |
| D | Step graded GeSn | 16.05 | -0.80 | 0.053 | ~ 0* | 0.146 |

TABLE I: Layer properties and band structure parameters calculated at 300 K for micro-disks before and after strain relaxation. (*) Strain in partially suspended micro-disks measured by Raman spectroscopy (Supplementary Information).

We evaluated the optical quality of our layers and the directness of the bandgaps through temperature-dependent photoluminescence (PL) (see Supplementary Information). Figure 1a and 1b show the integrated intensity and the Full Width at Half Maximum (FWHM) of the PL peaks including the luminescence from a 0.7 µm thick reference Ge layer [30] extracted from the spectra provided in Figures S3a-d in Supplementary Information. When the measurement temperature decreases from 300 K to 20K, the $\Gamma$-valley luminescence of all GeSn samples strongly increases. This trend is a clear indication that the bandgap is direct [3]. By contrast, the PL intensity of the Ge on Si sample, with therefore an indirect bandgap, is much lower and decreases when the sample is cooled-down. Interestingly, the integrated PL intensity of sample C (GeSn 13 % layer on a GeSn step-graded buffer) increases significantly only for temperatures below 150-200 K. This is likely due to the very small energy difference between the L and $\Gamma$ valleys, i.e. only 15 meV (Table I), allowing for thermal escape of electrons from the $\Gamma$ conduction band valley into the L valley at room temperature ($k_B T \sim 25$ meV). An additional proof of direct bandgap behavior is a significantly larger FWHM at room temperature for sample C: a shoulder in the PL spectra is indeed observed (Figure S3), which is likely coming from the L-valley.

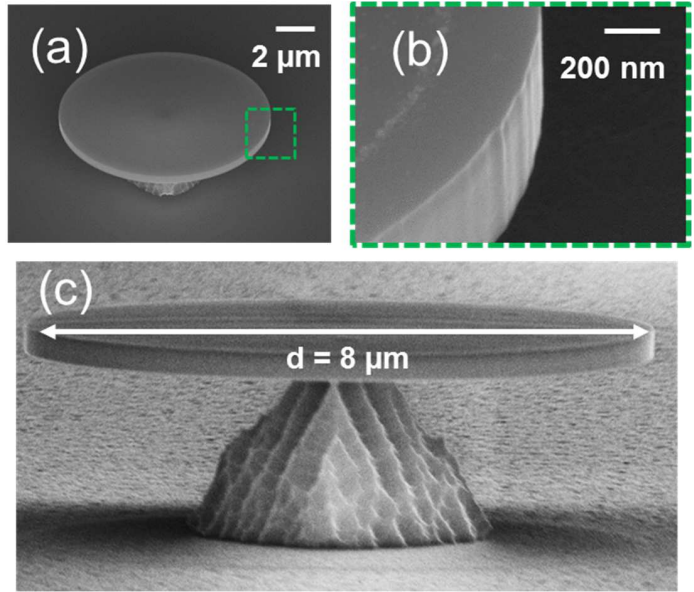

Figure 2: (a) 45° tilted view SEM image of a GeSn micro-disk; (b) zoom on the etched sidewall for the same 45° tilt and (c) a 5° tilt view of that micro-disk.

Using the GeSn layers described above as the active media, we fabricated suspended micro-disk optical resonators [6,27]. This approach enables to (i) increase the offset between Γ and L valleys of the active GeSn layers by relaxing the strain (Table I) and (ii) confine optical modes in the GeSn layers. The layers were patterned using e-beam lithography and an anisotropic dry etching with $Cl_2/N_2$ gasses. The GeSn layers were then under-etched using a selective dry etching recipe based on $CF_4$ [31,32]. The dry etching recipe selectively etched the Ge SRBs and the 8% Sn layers, leaving intact the higher Sn content GeSn layers above. This way, most of the misfit dislocations at the Ge/GeSn interfaces were removed. The thickness of sample C (Sn 13% on step graded GeSn buffer) is then 305 nm, to be compared to ~ 450 nm for sample A (Figure S3 inset). The 400 nm thickness of sample D (Sn 13% on step graded GeSn buffer) is by contrast close to the 430 nm of sample B (Figure S3 inset). The surface of the micro-disks was not passivated. Figure 2 shows tilted Scanning Electron Microscopy (SEM) views of a typical micro-disk with an 8 µm diameter, smooth sidewalls (Figure 2a and b) and a 3.8 µm under-etch (Figure 2c).

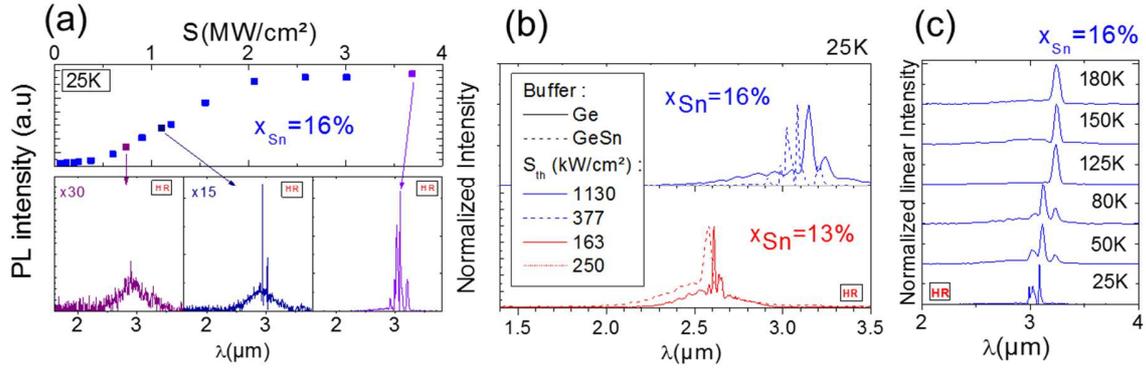

Figure 3: GeSn micro-disks characterization: (a) L-L curve of a 8µm micro-disk fabricated in the 16 % GeSn layer grown on the GeSn graded buffer at 25 K (sample D); (b) Laser spectra for all samples at 25 K with the laser threshold indicated in the legend; (c) Evolution of the laser spectra as a function of temperature (spectra from 50 K to 150K acquired with a low spectral resolution).

To test the lasing performances of the micro-disks, we used a 1064 nm Nd:YAG pulsed pump laser with a 0.6 ns pulse width and a 50 kHz repetition rate. The pump was focused on micro-disks with a spot diameter of 8 µm. Figure 3 shows the laser-output versus pumping-laser-input (L-L) for a 8µm micro-disk fabricated in the GeSn layer with 16 % Sn content grown on the GeSn step-graded buffer (i.e. sample D), together with emission spectra at three different pumping levels at 25 K. The lasing threshold is measured at 377 kW/cm² at 25 K. Figure 3b presents the lasing spectra of all samples at 25 K at their corresponding threshold power. The emitted wavelength is close to 2.6 µm for the $Ge_{0.87}Sn_{0.13}$ lasers and around 3.1 µm for the $Ge_{0.84}Sn_{0.16}$ lasers. The laser thresholds deduced from measured spectra and the detected power are indicated in the caption. For the $Ge_{0.87}Sn_{0.13}$ lasers, the laser thresholds were equal to 163 kW/cm² with the Ge buffer (sample A) and 250 kW/cm² for the step-graded GeSn buffer (sample C) at 25 K, i.e. values close to those reported in passivated micro-disks from the Forschungszentrum Juelich [6]. The higher laser thresholds of sample C can be explained by the lower thickness of the GeSn layers leading to a lower mode confinement for sample C (435 nm for sample A, to be compared to 305 nm only for sample C). By contrast, the situation is the opposite for 16% of Sn when using a step-graded GeSn buffer instead of a Ge SRB, as illustrated in Figure 3b. The laser threshold for sample D was indeed at around 377 kW/cm², to be compared with 1130 kW/cm² for sample B, i.e. a threshold nearly three times lower. This can be attributed to a better material quality as the mode confinement is expected to be similar (430 nm for sample B, 400 nm for sample D) [22]. Figure 3c shows laser spectra as a function of temperature for sample D. The spectra from 50 K to 180 K were acquired with a lower spectral resolution. The 180 K spectra (Figure 3c) have been measured at 4.3MW/cm². As the temperature increases from 50 K to 180 K, a mode switching was observed. Indeed, the optical gain shifted rapidly with the temperature, while the cavity modes evolved at a much slower rate.

At wavelengths around 3.1 µm, we observed lasing operation up to 180 K. We, thus, significantly improved the operating temperature over the previously reported values of 110 K [7] and 130 K [6]. We attribute the improvement to a larger energy splitting between Γ- and L- valleys in higher Sn-content layers (16% in our case compared to 12.5 % in refs [3,6,7]). We expect the thresholds of our lasers to be reduced using surface passivation and carrier confinement. In addition, capping of the top GeSn layers with lower Sn content layers (i.e. the use of double heterostructures) should enhance carrier confinement.

In summary, we have studied partially relaxed GeSn layers with very high Sn contents (up to 16%) grown either on Ge SRBs or on step-graded GeSn buffers. Both buffer strategies led to a direct bandgap for $Ge_{0.87}Sn_{0.13}$ and $Ge_{0.84}Sn_{0.16}$ layers. After under-etching, both buffer strategies yielded optically pumped laser emission at 2.6 µm (for 13% of Sn) and 3.1 µm (for 16% of Sn). However, only GeSn layers grown on step-graded buffers presented mirror-like surfaces under grazing light with no significant Sn segregation over the whole 200-mm wafer surface and only a combination of 16% Sn GeSn layers with step-graded GeSn buffers led to a laser operation up to 180 K. The Ge$_{0.84}$Sn$_{0.16}$ microdisk cavities emitted at 3.1 µm with a lasing threshold of 377 kW/cm² at 25K. An increase of Sn content grown on GeSn step-graded buffers is, thus, a path towards room-temperature GeSn lasers.

**Acknowledgements**

The authors would like to thank the "Plateforme de Technologie Amont" and the "41" platform of the CEA Grenoble for the clean room facilities and acknowledge TEEM Photonics for the loan of the pump laser. We thank Eugénie Martinez for her help in characterizing GeSn samples. This work was supported by the CEA DRF-DRT Phare Photonics and IBEA Nanoscience projects as well as the Swiss National Science foundation SN

**Supporting information:** Optically pumped GeSn micro-disks with 16 % Sn lasing at 3.1 µm up to 180K, V. Reboud[*], A. Gassenq, N. Pauc, J. Aubin, L. Milord, Q. M. Thai, M. Bertrand, K.Guilloy, D. Rouchon, J. Rothman, T. Zabel, F. Armand Pilon, H. Sigg, A. Chelnokov, J.M. Hartmann, V. Calvo.

1. <u>XRD characterizations of GeSn layers</u>

The GeSn layers of samples A and B were grown directly on 2.5 µm thick Ge SRBs. Meanwhile, the GeSn step-graded structures of samples C and D were grown on 1.3 µm thick Ge SRBs. Figure S1 a-d shows the reciprocal space maps (RSM) around the (224) asymmetric diffraction order of those samples. Measured layer thicknesses and Sn concentrations are reported in the inset of Figure S1 and in [33] for GeSn with 16% of Sn. For Samples A and B, two distinct GeSn peaks are detected on each profile (Figures S1a and S1b). Those peaks are associated with the unexpected presence of two Sn concentrations in the GeSn layers grown on Ge SRBs. Such a trend was confirmed by TEM-EDX measurements and was due to plastic strain relaxation in such thick layers [33]. As expected, three and four intense peaks coming from the various GeSn layers, were observed for samples C and D, respectively.

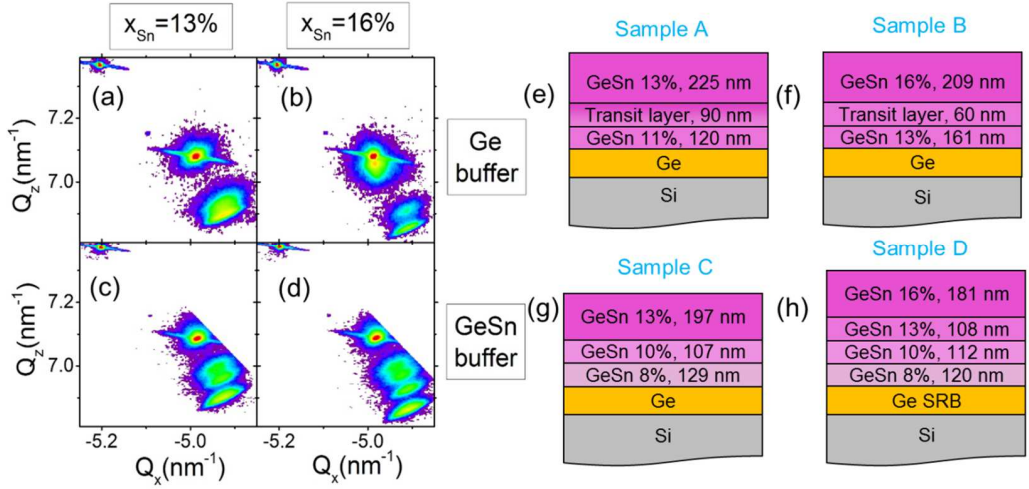

Figure S1: (a-d): RSM measurements of GeSn layers with 13% and 16% of Sn grown directly on Ge or on GeSn step graded buffers on Ge SRBs: (a) sample A (Sn 13% on Ge SRB); (b) sample B (Sn 16% on Ge SRB); (c) sample C (Sn 13% on step graded GeSn buffer) and (d) sample D (Sn 16% on step-graded GeSn buffer). Schematic representation of (e,f) thick, nominally constant composition $Ge_{1-x}Sn_x$ active layers grown on 2.5 µm thick Ge SRBs and (g,h) GeSn step graded buffers grown on 1.3 µm thick Ge SRBs.

|        |                  | Before strain relaxation | | | | |
| ------ | ---------------- | ---- | ------------ | ------- | ------- | ------------- |
| Sample | Growth buffer    | $x_{Sn}$ | $\varepsilon_{//}$ | $Eg_\Gamma$ | $Eg_L$ | $Eg_L - Eg_\Gamma$ |
|        |                  | %    | %            | eV      | eV      | eV            |
| A      | Ge SRB           | 13.07 | -0.56       | 0.449   | 0.479   | .030          |
| B      | Ge SRB           | 16.41 | -0.76       | 0.379   | 0.442   | .063          |
| C      | Step graded GeSn | 12.87 | -0.66       | 0.469   | 0.484   | .015          |
| D      | Step graded GeSn | 16.05 | -0.80       | 0.394   | 0.447   | .053          |

TABLE II: Detailed layers properties before strain relaxation and their associated band structure parameters calculated at 300 K.

2. Strain characterization via Raman spectroscopy

We performed Raman spectroscopy measurements on fabricated micro-disks. A Renishaw InVia Raman spectrometer with a resolution of ±0.6 cm$^{-1}$ was used together with a 532 nm incident laser with a ~20 nm penetration depth [34,35] and a 0.7 µm-diameter spot. Figures S2a-b show the Raman spectra for micro-disks fabricated with Sn concentration of 13 and 16% as well as with Ge and GeSn buffers. The Raman spectral shift was measured by fitting the spectra with Lorentzian functions and comparing their wavenumbers to that of a bulk Ge substrate. The Raman spectral shifts before processing were of 8.8 cm$^{-1}$ and 11.3 cm$^{-1}$ for Ge$_{0.87}$Sn$_{0.13}$ and Ge$_{0.84}$Sn$_{0.16}$ micro-disks, respectively, corresponding to a residual compressive strain of 0.5 % [35]. This value is in good agreement with the strain measured by XRD (table I). After under-etching, the Raman spectral shift increased up to 11.6 cm$^{-1}$ and up to 13.6 cm$^{-1}$ for 13% Sn and 16% Sn micro-disks, respectively, indicating that the residual strain was completely relaxed [35]. The directness of the active layers is then expected to increase as the Γ and L valley split is amplified by strain relaxation [Table I].

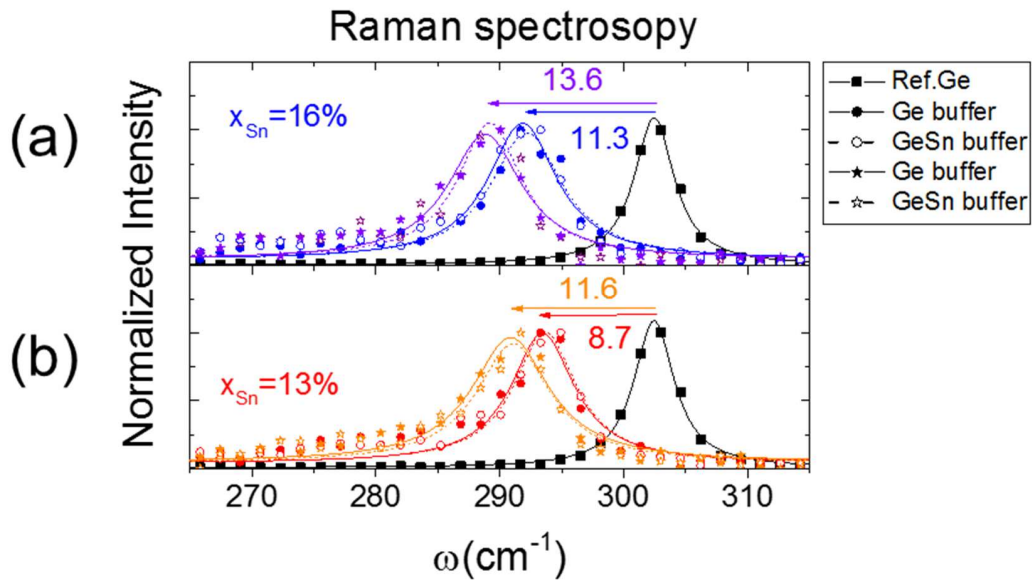

Figure S2: Raman spectroscopy measurements performed, before and after under etching, on (a) Ge$_{0.84}$Sn$_{0.16}$ micro-disks directly grown on Ge SRBs and on GeSn step-graded buffers and (b) Ge$_{0.87}$Sn$_{0.13}$ micro-disks grown on Ge SRB and on GeSn step-graded buffers.

3. Photoluminescence characterization of GeSn layers

PL was excited with a 1047 nm continuous wave laser. The laser light was focused on a 20 µm diameter spot with a 1 mW average power. The light emitted by the samples was analyzed with a Fourier Transform Infra-Red spectrometer [36] equipped with a mercury cadmium telluride avalanche photodetector [37]. For each of the four samples, PL spectra as a function of the temperature are shown in Figures S3a to S3d. Log scales are used for a better reading. As expected, the Γ bandgap decreases as the Sn content increases [38,39,40].

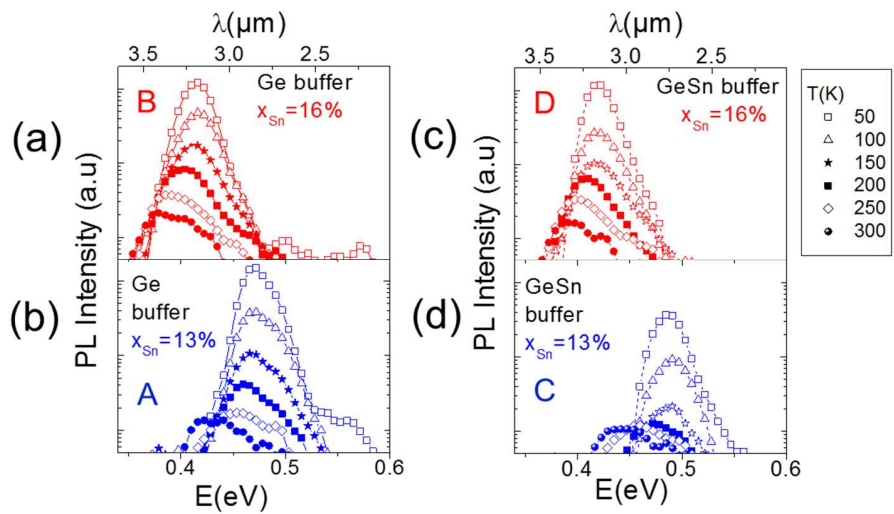

Figure S3: Temperature-dependent photoluminescence (PL) measurements: PL spectra for samples A and B, i.e. GeSn layers grown directly on Ge SRBs with (a) 16 % and (b) 13 % of Sn or for samples C and D, i.e. thick GeSn layers grown on GeSn step-graded buffers with (c) 16% and (d) 13 % of Sn.